\documentclass[12pt]{article}
\usepackage{graphicx,cite,amsmath,epsfig}    
\usepackage{amssymb}
\usepackage{amsmath,color} 
\usepackage{url}
\usepackage{xcolor}
\usepackage{epsfig}
\usepackage{footnote}
\usepackage{multirow}
\usepackage{enumitem}
\graphicspath{{figs/}}
\usepackage{float}
\usepackage{color}    
\usepackage{cite}
\usepackage{mathtools}

\def\beq{\begin{equation}}
\def\eeq{\end{equation}}
\def\nn{\nonumber}
\def\bea{\begin{eqnarray}}
\def\eea{\end{eqnarray}}
\def\ba{\begin{array}}                  
\def\ea{\end{array}}

\newcommand{\vt}{\vert}
\newcommand{\mc}{\mathcal}
\def\dg{\dagger}     

\include{paperdef}

\evensidemargin \oddsidemargin
\usepackage[latin1]{inputenc}

\begin{document}

\thispagestyle{empty}

\def\thefootnote{\fnsymbol{footnote}}


\vspace{1cm}

\begin{center}
{\Large \sc {\bf B-L Model with  ${\bf S}_{3}$ Symmetry \\ 
\normalsize Nearest Neighbor Interaction Textures and Broken $\mu\leftrightarrow\tau$ Symmetry}}

\vspace{0.4cm}

\vspace{1cm}

{
\sc Juan Carlos G\'omez-Izquierdo$^{1,2,3}$\footnote{email: jcgizquierdo1979@gmail.com}, Myriam Mondrag\'on$^{2}$\footnote{email: myriam@fisica.unam.mx}
}

\vspace*{1cm}

{\sl $^{1}$
Centro de Estudios Cient\'ificos y Tecnol\'ogicos No 16, Instituto Polit\'ecnico Nacional, Pachuca: Ciudad del Conocimiento y la Cultura, Carretera Pachuca Actopan km 1+500, San Agust\'in Tlaxiaca, Hidalgo, M\'exico.\\
\sl $^{2}$ Instituto~de F{\'{\i}}sica, Universidad~Nacional Aut\'onoma de M\'exico, 
M\'exico 01000, D.F., M\'exico.\\
\sl $^{3}$ Departamento de F\'isica, Centro de Investigaci\'on y de Estudios
Avanzados del I. P. N.,\\
Apdo. Post. 14-740, 07000, Ciudad de M\'exico, M\'exico.\\
}

\end{center}
\vspace*{0.2cm}
\begin{abstract}
We make a scalar extension of the B-L gauge model where the
${\bf S}_{3}$ non-abelian discrete group drives mainly the Yukawa
sector. Motived by the large and small hierarchies among the quark and
active neutrino masses respectively, the quark and lepton families
are not treated on the same footing under the assignment of the
discrete group. As a consequence, the Nearest Neighbor Interactions
(NNI) textures appear in the quark sector, leading to the CKM mixing matrix, whereas in the lepton
sector, a soft breaking of the $\mu \leftrightarrow \tau$ symmetry 
in the effective neutrino mass that comes from type I see-saw
mechanism, provides a non-maximal atmospheric angle and a non-zero
reactor angle.
\end{abstract}

\def\thefootnote{\arabic{footnote}}
\setcounter{page}{0}
\setcounter{footnote}{0}
\newpage

\section{Introduction}
How to explain and understand tiny neutrino masses and the fermion
mixings respectively in and beyond the Standard Model (SM) is still an 
open question. Up to now, it is not clear if there is an organizing
principle in the Yukawa sector that explains the almost diagonal CKM
mixing matrix and its counterpart PMNS one, that has large mixing values.

The pronounced hierarchy among the quark masses,
$m_{t}\gg m_{c}\gg m_{u}$ and $m_{b}\gg m_{s}\gg m_{d}$, could be
behind the small mixing angles that parametrize the CKM, which
depend strongly on the mass ratios \cite{Fritzsch:1999ee,
Xing:2014sja, Verma:2015mgd}. From a phenomenological point of view,
hierarchy among the fermion masses may be understood by means of
textures (zeros) in the fermion mass matrices \cite{Fritzsch:1999ee,
Xing:2014sja, Verma:2015mgd, Tanimoto:2016rqy}. On the theoretical
side, mass textures can be generated dynamically by non-abelian
discrete symmetries \cite{Ishimori:2010au, Altarelli:2012bn,
Grimus:2011fk, Altarelli:2012ss, King:2013eh}. The Fritzsch
\cite{Fritzsch:1977vd, Fritzsch:1979zq, Fritzsch:1985eg} and the NNI
\cite{Branco:1988iq, Branco:1994jx, Harayama:1996am, Harayama:1996jr}
textures are hierarchical, however, only the latter one can
accommodate with good accuracy the CKM matrix.

In the lepton sector, the hierarchy seems to work differently in the mixings
since the charged lepton masses are hierarchical,
$m_{\tau}\gg m_{\mu}\gg m_{e}$, but the active neutrino masses
exhibit a weak hierarchy \cite{Fritzsch:2009sm, Fritzsch:2015foa} that
may be responsible for the large mixing values. If the
neutrinos obey a normal mass ordering, large mixings can also be 
obtained by the Fritzsch and NNI textures \cite{Fritzsch:2009sm,
Fritzsch:2011cu, Fritzsch:2015foa}. It is worth mentioning that
non-hierarchical fermion mass matrices could also accommodate the
lepton mixing angles \cite{Ludl:2014axa, Fritzsch:2015gxa}. Nevertheless, the hierarchy might have nothing to do
with the mixing \cite{Holthausen:2012dk, Holthausen:2012wt, Fonseca:2014koa, Joshipura:2016quv}, since large mixings might be explained by discrete
symmetries which were motivated mainly by the experimental values,
$\theta_{23}\approx 45^{\circ}$ and $\theta_{13}\approx0^{\circ}$. The
$\mu \leftrightarrow \tau$ symmetry \cite{Fukuyama:1997ky,
Mohapatra:1998ka, Lam:2001fb, Kitabayashi:2002jd, Grimus:2003kq,
Koide:2003rx} was proposed to be behind the atmospheric and reactor
mixing angle values. This symmetry predicts exactly that
$\theta_{23}=45^{\circ}$ and $\theta_{13}=0^{\circ}$, which were
consistent with the experimental data many years ago. The
Tri-Bimaximal (TB) mixing pattern \cite{Harrison2002167, Xing200285,
Altarelli:2012ss} was suggested for obtaining the above angles
plus $\sin{\theta_{12}}={1/\sqrt{3}}$, for the solar angle which
coincides approximately with the experimental value.
An intriguing fact is that the above mixing pattern does not depend on
the lepton masses up to corrections to the lower order in the mixing
matrices. 

To face the neutrino masses and mixings problems, one
has to go beyond the SM, and it has to be extended or
replaced by a new framework where ideally both issues can be
explained. In this line of thought, one of the best
motivated candidates to replace the SM is the baryon number minus
lepton number (B-L) gauge model, $SM\otimes U(1)_{B-L}$, which may
come from the Grand Unified Theory (GUT) $SO(10)$
\cite{Fritzsch:1974nn, Buchmuller:1991ce} or from the unified
model $S(3)_{C}\otimes SU(3)_{L}\otimes U(1)_{X}\otimes U(1)_{N}$
\cite{Dong:2015yra, Dong:2017zxo}. The breaking mass scale of the
B-L model to the SM is related with the mass of the three
right-handed neutrinos (RHN's) that are included to cancel
anomalies and explaining, at tree level, the tiny neutrino masses
by means of the type I see-saw mechanism \cite{Minkowski:1977sc,
GellMann:1980vs, Yanagida:1979as, Mohapatra:1979ia,
Schechter:1980gr, Mohapatra:1980yp, Schechter:1981cv} (for other
mechanism in B-L see \cite{Perez:2017qns}). Aparte from neutrino
masses and mixings, leptogenesis, dark matter and inflation also have
found a realization in the B-L model \cite{Khalil:2006yi,
Emam:2007dy, Abbas:2007ag, Khalil:2008ps, Higaki:2014dwa,
Guo:2015lxa, Dev:2017xry}. Due to all those features, from our
point of view, the renormalizable B-L model has the main ingredients
to address the problem of the quark and lepton masses and their contrasting mixing
matrices.

Moving on to the mixing, the ${\bf S}_{3}$
  non-abelian group has been proposed as the underlying flavor
  symmetry in different frameworks \cite{Pakvasa:1977in, Kubo:2003iw,
    Kubo:2005sr, Mondragon:2006hi, Felix:2006pn, Mondragon:2007af,
    Mondragon:2007nk, Mondragon:2007jx, Meloni:2010aw, Dicus:2010iq,
    Canales:2011ug, Canales:2012ix, Canales:2012dr,
    GonzalezCanales:2012za, Canales:2013ura, Canales:2013cga,
    Hernandez:2013hea, Hernandez:2014lpa, Ma:2014qra,
    Hernandez:2014vta, Hernandez:2015dga, Hernandez:2015zeh,
    Das:2015sca, Hernandez:2015hrt, Pramanick:2016mdp,
    Arbelaez:2016mhg, CarcamoHernandez:2016pdu,
    CarcamoHernandez:2017cwi, Cruz:2017add, Ge:2018ofp}. One
  motivation to use this discrete symmetry in the lepton sector is to
  generate the $\mu\leftrightarrow\tau$ symmetry
  ~\cite{Mohapatra:1998ka, Lam:2001fb, Kitabayashi:2002jd,
    Grimus:2003kq, Koide:2003rx, Fukuyama:1997ky} or the TB mixing
  matrix \cite{Harrison:2003aw, Mohapatra:2006pu}.  In the quark
  sector, the ${\bf S}_{3}$ symmetry can give rise to the Fritzsch and
  generalized Fritzsch mass textures \cite{Meloni:2010aw,
    Barranco:2010we}. More recently, it was shown
    that the NNI mass textures are hidden in the ${\bf S}_{3}$ flavor
    symmetry \cite{Canales:2013cga}, this last novel fact will be
    highlighted as  part of our motivation in the present work.

Therefore, we make a scalar extension of the B-L gauge model where the
${\bf S}_{3}$ non-abelian discrete group drives mainly the Yukawa
sector. Motived by the large and small hierarchies among the quark and
active neutrino masses respectively, the quark and lepton families are
not treated on the same footing under the assignment of the discrete
group. As a consequence, NNI textures appear in the quark sector,
leading to the CKM mixing matrix, whereas in the lepton sector, a soft
breaking of the $\mu \leftrightarrow \tau$ symmetry in the effective
neutrino mass that comes from type I see-saw mechanism, provides a
non-maximal atmospheric angle and a non-zero reactor angle.

The plan of this paper is as follows: the B-L gauge model and the
${\bf S}_{3}$ flavor symmetry are described briefly in Section 2, the
fermion masses and mixings will be discussed in Section 3. In section
4 some conclusions are drawn.

\section{Flavored B-L Model}
The B-L gauge model is based on the $SU(3)_{c}\otimes SU(2)_{L}\otimes
U(1)_{Y}\otimes U(1)_{B-L}$ gauge group where, apart from the SM
fields, three $N_{i}$ RHN's and a $\phi$ singlet scalar field are
added to the matter content. Under B-L, the quantum numbers for
quarks, leptons and Higgs ($\phi$) are $1/3$, $-1$ and $0$ ($-2$),
respectively. The allowed Lagrangian is
\begin{equation}
\mathcal{L}_{B-L}=\mathcal{L}_{SM}-y^{D}\bar{L}\tilde{H}N-\frac{1}{2}y^{N}\bar{N}^{c}\phi N-V\left(H,\phi\right)
\end{equation}
with
\begin{equation}
V\left(H,\phi \right)=\mu^{2}_{BL}\phi^{\dagger} \phi+\frac{\lambda_{BL}}{2}\left(\phi^{\dagger}\phi\right)^{2}-\lambda^{H \phi}\left(H^{\dagger} H \right)\left(\phi^{\dagger} \phi \right).\label{sp1}
\end{equation}
where $\tilde{H}_{i}=i\sigma_{2}H^{\ast}_{i}$. The spontaneous
symmetry breaking of $U(1)_{B-L}$ happens usually at high energies so
the breaking scale is larger than the electroweak scale, $\phi_{0}\gg
v$. In this first stage, the RHN's become massive particles, along
with this, an extra gauge boson, $Z_{B-L}$, appears as a result of
breaking the gauge group. The rest of the particles turn out massive
when the Higgs scalars acquire their vacuum expectation value (vev's) and
tiny active neutrino masses are explained by the type I see-saw
mechanism.
\begin{align}
\langle H\rangle= \frac{1}{\sqrt{2}}\left(
\ba{c}
0 \\
v \\
\ea
\right), \quad \langle \phi\rangle=\frac{\phi_{0}}{\sqrt{2}}.\label{eq6.1}
\end{align}

On the other hand, let us describe briefly the non-Abelian group ${\bf
S}_{3}$, which is the permutation group of three objects; it 
has three irreducible representations: two 1-dimensional, ${\bf
1}_{S}$ and ${\bf 1}_{A}$, and one 2-dimensional representation,
${\bf 2}$ (for a detailed study see \cite{Ishimori:2010au}). So, the
three dimensional real representation can be decomposed as: ${\bf
3}_{S}={\bf 2}\oplus {\bf 1}_{S}$ or ${\bf 3}_{A}={\bf 2}\oplus {\bf
1}_{A}$. The multiplication rules among the irreducible
representations are
\begin{align}\label{rules}
&{\bf 1}_{S}\otimes {\bf 1}_{S}={\bf 1}_{S},\quad {\bf 1}_{S}\otimes {\bf 1}_{A}={\bf 1}_{A},\quad {\bf 1}_{A}\otimes {\bf 1}_{S}={\bf 1}_{A},\quad {\bf 1}_{A}\otimes {\bf 1}_{A}={\bf 1}_{S},\nn\\&
{\bf 1}_{S}\otimes {\bf 2}={\bf 2},\quad {\bf 1}_{A}\otimes {\bf 2}={\bf 2},\quad {\bf 2}\otimes {\bf 1}_{S}={\bf 2},\quad {\bf 2}\otimes {\bf 1}_{A}={\bf 2};\nn\\
&\begin{pmatrix}
a_{1} \\ 
a_{2}
\end{pmatrix}_{{\bf 2}}
\otimes
\begin{pmatrix}
b_{1} \\ 
b_{2}
\end{pmatrix}_{{\bf 2}} = 
\left(a_{1}b_{1}+a_{2}b_{2}\right)_{{\bf 1}_{S}} \oplus  \left(a_{1}b_{2}-a_{2}b_{1}\right)_{{\bf 1}_{A}} \oplus	
\begin{pmatrix}
a_{1}b_{2}+a_{2}b_{1} \\ 
a_{1}b_{1}-a_{2}b_{2}
\end{pmatrix}_{{\bf 2}}. 
\end{align}

Having introduced the theoretical framework and the flavor symmetry
that will play an important role in the present work, it is worthwhile
to point out that the present idea has been developed in the framework
of left-right symmetric model (LRSM)
\cite{Gomez-Izquierdo:2017rxi}. However, there are substantial
differences between the LRSM and B-L models in their minimal versions:
a) due to the gauge symmetry the LRSM contains more scalars fields in
comparison to the B-L model, b) as a consequence the latter one has a
simpler Yukawa mass term than the LRSM, which allows us to work with
fewer couplings in the fermionic mass matrices; c) the effective
neutrino mass matrix has two contributions due to the type I and II
see-saw mechanism in the LRSM (the latter one usually is neglected by
hand) whereas the type I see-saw mechanism works out in the B-L model. The
above statements stand for some advantages to study fermion masses and
mixings in the B-L model, moreover, the LRSM comes out being
complicated if the scalar sector is augmented. From
  these comments, one may conclude that the current work is a simple
  comparison with the LRSM, however, we emphasize that the quark
  sector makes the difference between the present work and that
  developed in \cite{Gomez-Izquierdo:2017rxi}. As we will see later,
  in the current study the CKM matrix is understood by hierarchical
  mass matrices which is not the case in
  \cite{Gomez-Izquierdo:2017rxi}; this last statement can be verified in
  \cite{Garces:2018nar} which is an extended version of the previous
  work\cite{Gomez-Izquierdo:2017rxi}.

Now, let us remark important points about the scalar sector and the
family assignment under the flavor symmetry in our model. Due to the
flavor symmetry, three Higgs doublets have been added in this model to
obtain the CKM mixing matrix. At the same time, as we will see, the
charged leptons and Dirac neutrinos mass matrices are built to be
diagonal, so the mixing will come from the RHN mass matrix. Then, three
singlets scalars fields, $\phi_{i}$, are needed to accomplish
this. Along with this, in order to try  explaining naively the
contrasting values between the CKM and PMNS mixing matrices, let us point
out a crucial difference in the way the quark and lepton families have
been assigned under the irreducible representations
of ${\bf S}_{3}$. Hierarchy among the fermion masses makes suggests 
that both in the quark and Higgs sector, the first
and second family are put together in a flavor doublet ${\bf
2}$ and the third family in a singlet ${\bf 1}_{S}$. On the 
contrary, for the leptons, the first family has been assigned to a singlet ${\bf 1}_{S}$
and the second and third families to  a doublet ${\bf 2}$. As
consequence of this assignment, the hierarchical NNI textures are
hidden in the quark mass matrices. In the lepton sector, on the other
hand, the lepton mixings can be understood from an approximated $\mu
\leftrightarrow \tau$ symmetry in the effective neutrino mass matrix
\cite{Gomez-Izquierdo:2017rxi}.

We ought to comment that the above flavor symmetry
  assignment may be incompatible with $SO(10)$ multiplets, however,
  this assignment could be realized in the
  $S(3)_{C}\otimes SU(3)_{L}\otimes U(1)_{X}\otimes U(1)_{N}$ model (see for instance \cite{Dong:2015yra, Dong:2017zxo}).

In Table \ref{tab2}, the full assignment for the matter content is
shown. The ${\bf Z_{2}}$ symmetry has been added in order to prohibit
some Yukawa couplings in the lepton sector, but this is not enough to
obtain diagonal mass matrices. So, an extra symmetry will be imposed
below.
\begin{table}[ht]
\begin{center}
\begin{tabular}{|c|c|c|c|c|c|c|c|c|c|c|c|c|c|c|}
\hline \hline	
{\footnotesize Matter} & {\footnotesize $Q_{I L}, H_{I}, d_{I R}, u_{I R}$, $\phi_{I}$} & {\footnotesize $Q_{3 L}, H_{3}, d_{3 R}, u_{3 R}$, $\phi_{3}$} & {\footnotesize $L_{1}, e_{1 R}, N_{1}$} & {\footnotesize $L_{J}, e_{J R}, N_{J}$}   \\ \hline
{\footnotesize \bf $S_{3}$} &  {\footnotesize \bf $2$} & {\footnotesize \bf $1_{S}$}   & {\footnotesize \bf $1_{S}$} & {\footnotesize \bf $2$} \\ \hline
{\footnotesize \bf $Z_{2}$} & {\footnotesize $1$} & {\footnotesize $1$}  &  {\footnotesize $1$} & {\footnotesize $-1$} \\ \hline \hline
\end{tabular}\caption{Flavored $B-L$ model. Here, $I=1,2$ and $J=2,3$.}\label{tab2}
\end{center}
\end{table}
Thus, the most general form for the Yukawa interaction Lagrangian that respects the ${\bf S}_{3}\otimes {\bf Z}_{2}$ flavor symmetry and the gauge group, is given as
\begin{align}
-\mathcal{L}_{Y}&=y^{d}_{1}\left[\bar{Q}_{1 L}\left(H_{1} d_{2 R}+H_{2} d_{1 R}\right)+\bar{Q}_{2 L}\left(H_{1} d_{1 R}-H_{2}d_{2 R}\right)\right]+y^{d}_{2}\left[\bar{Q}_{1 L}H_{3}d_{1 R}+\bar{Q}_{2 L}H_{3} d_{2 R}\right]\nn\\&+y^{d}_{3}\left[\bar{Q}_{1 L}H_{1}+\bar{Q}_{2 L}H_{2}\right]d_{3 R}+y^{d}_{4}\bar{Q}_{3 L}\left[H_{1}d_{1 R}+H_{2}d_{2 R}\right]+y^{d}_{5}\bar{Q}_{3 L}H_{3}d_{3 R}\nn\\&+y^{u}_{i}
\left(H\rightarrow \tilde{H},d_{R}\rightarrow u_{R}\right)
+y^{e}_{1}\bar{L}_{1}H_{3}e_{1 R}+y^{e}_{2}\left[(\bar{L}_{2}H_{2}+\bar{L}_{3}H_{1})e_{2 R}+(\bar{L}_{2}H_{1}-\bar{L}_{3}H_{2})e_{3 R} \right]\nn\\&+y^{e}_{3}\left[\bar{L}_{2}H_{3}e_{2 R}+\bar{L}_{3}H_{3}e_{3 R}\right]+
y^{D}_{i}\left(H\rightarrow \tilde{H},e_{R}\rightarrow N\right)
+y^{N}_{1}\bar{N}^{c}_{1}\phi_{3}N_{1}\nn\\&+y^{N}_{2}\left[\bar{N}^{c}_{1}\left(\phi_{1 }N_{2}+\phi_{2}N_{3}\right)+\left(\bar{N}^{c}_{2}\phi_{1}+\bar{N}^{c}_{3}\phi_{2}\right)N_{1}\right]+y^{N}_{3}\left[\bar{N}^{c}_{2}\phi_{3 }N_{2}+\bar{N}^{c}_{3}\phi_{3}N_{3}\right]+h.c.\label{eq2}
\end{align}
At this stage, an extra symmetry ${\bf Z}^{e}_{2}$ is used to obtain
diagonal charged and neutrinos Dirac mass matrices. This symmetry does
not modify the Majorana mass matrix form. Explicitly, in the above
Lagrangian, we demand that
\begin{align}
L_{3}\leftrightarrow-L_{3},\quad e_{3 R}\leftrightarrow-e_{3 R},\quad  N_{3 }\leftrightarrow-N_{3},\quad \phi_{2 }\leftrightarrow -\phi_{2}.\label{exss}
\end{align}
so the off-diagonal entries $23$ and $32$ in the lepton sector are
absent. Then, this allows to identify properly the charged lepton
masses, at the same time, we can speak strictly about the $\mu
\leftrightarrow \tau$ symmetry in the effective neutrino mass.

On the other hand, it is convenient to point out that the scalar potential of the SM with three families of
Higgs, $V(H_{i})$, and the representation ${\bf 3}_{S}={\bf 2}\oplus {\bf 1}_{S}$
has been studied in \cite{Pakvasa:1977in, Kubo:2004ps, EmmanuelCosta:2007zz, Beltran:2009zz, Bhattacharyya:2012ze, Teshima:2012cg, Das:2014fea,
Barradas-Guevara:2014yoa}. So, in the $B-L$
model, the flavored gauge scalar potential (together with
Eq. (\ref{exss})) is given by
$V(H_{i},\phi_{i})=V(H_{i})+V(\phi_{i})+V(H_{i},\phi_{i})$ where the
first term has already been analyzed in the mentioned works; the second and third terms are given as 
{\scriptsize
\begin{align}
V(\phi_{i}) +V(H_{i},\phi_{i})&=\mu^{2}_{1BL}\left(\phi^{\dagger}_{1}\phi_{1}+\phi^{\dagger}_{2}\phi_{2}\right)+\mu^{2}_{2BL}\left(\phi^{\dagger}_{3}\phi_{3}\right) +\lambda^{\phi}_{1}\left(\phi^{\dagger}_{1}\phi_{1}+\phi^{\dagger}_{2}\phi_{2}\right)^{2}+\lambda^{\phi}_{2}\left(\phi^{\dagger}_{1}\phi_{2}-\phi^{\dagger}_{2}\phi_{1}\right)^{2}\nn\\&+\lambda^{\phi}_{5}\left(\phi^{\dagger}_{3}\phi_{3}\right)\left(\phi^{\dagger}_{1}\phi_{1}+\phi^{\dagger}_{2}\phi_{2}\right)
+\lambda^{\phi}_{3}\left[\left(\phi^{\dagger}_{1}\phi_{2}+\phi^{\dagger}_{2}\phi_{1}\right)^{2}+\left(\phi^{\dagger}_{1}\phi_{1}-\phi^{\dagger}_{2}\phi_{2}\right)^{2}\right]\nn\\&+\lambda^{\phi}_{6}\left[\left(\phi^{\dagger}_{3}\phi_{1}\right)\left(\phi^{\dagger}_{1}\phi_{3}\right)+\left(\phi^{\dagger}_{3}\phi_{2}\right)\left(\phi^{\dagger}_{2}\phi_{3}\right) \right]+\lambda^{\phi}_{7}\left[\left(\phi^{\dagger}_{3}\phi_{1}\right)^{2}+\left(\phi^{\dagger}_{3}\phi_{2}\right)^{2}+\textrm{h.c.}\right]+\lambda^{\phi}_{8}\left(\phi^{\dagger}_{3}\phi_{3}\right)^{2}\nn\\&+\lambda^{H\phi}_{1}\left(H^{\dagger}_{1}H_{1}+H^{\dagger}_{2}H_{2}\right) \left(\phi^{\dagger}_{1}\phi_{1}+\phi^{\dagger}_{2}\phi_{2}\right)+\lambda^{H\phi}_{4}\left(H^{\dagger}_{3}H_{2}\right) \left(\phi^{\dagger}_{1}\phi_{1}-\phi^{\dagger}_{2}\phi_{2}\right)+\lambda^{H \phi}_{5}\left(H^{\dagger}_{1}H_{2}+H^{\dagger}_{2}H_{1}\right) \left(\phi^{\dagger}_{3}\phi_{1}\right)\nn\\&+\lambda^{H \phi}_{6}\left(H^{\dagger}_{2}H_{3}\right) \left(\phi^{\dagger}_{1}\phi_{1}-\phi^{\dagger}_{2}\phi_{2}\right)+\lambda^{H \phi}_{7}\left(H^{\dagger}_{1}H_{2}+H^{\dagger}_{2}H_{1}\right) \left(\phi^{\dagger}_{1}\phi_{3}\right)+\lambda^{H \phi}_{8}\left(H^{\dagger}_{3}H_{3}\right)\left(\phi^{\dagger}_{1}\phi_{1}+\phi^{\dagger}_{2}\phi_{2}\right)\nn\\&+\lambda^{H \phi}_{9}\left(H^{\dagger}_{1}H_{1}+H^{\dagger}_{2}H_{2}\right)\left(\phi^{\dagger}_{3}\phi_{3}\right)+\lambda^{H \phi}_{10}\left(H^{\dagger}_{1}H_{3}\right) \left(\phi^{\dagger}_{3}\phi_{1}\right)+
\lambda^{H \phi}_{11}\left(H^{\dagger}_{3}H_{1}\right) \left(\phi^{\dagger}_{1}\phi_{3}\right)+\lambda^{H \phi}_{12}\left(H^{\dagger}_{3}H_{1}\right) \left(\phi^{\dagger}_{3}\phi_{1}\right)\nn\\&+\lambda^{H \phi}_{13}\left(H^{\dagger}_{1}H_{3}\right) \left(\phi^{\dagger}_{1}\phi_{3}\right)+\lambda^{H \phi}_{14}\left(H^{\dagger}_{3}H_{3}\right) \left(\phi^{\dagger}_{3}\phi_{3}\right).
\end{align}}
where the factor of $1/2$, in the second term of Eq. (\ref{sp1}), has been absorbed in the $\lambda^{\phi}_{i}\equiv\lambda_{BL}$ parameter. Then, assuming that all parameters in the scalar potential are real, the minimization condition for the complete scalar potential are give by
{\scriptsize
\begin{align}
2\mu^{2}_{1}&=-2\gamma\left(v^{2}_{1}+v^{2}_{2}\right)-6\lambda_{4}v_{2}v_{3}-\lambda v^{2}_{3}+\lambda^{H\phi}_{1}\left(\phi^{2}_{01}+\phi^{2}_{02}\right)+\left[\lambda^{H\phi}_{II}\frac{v_{2}}{v_{1}}+\lambda^{H\phi}_{III}\frac{v_{3}}{2v_{1}}
\right]\phi_{01}\phi_{03}+\lambda^{H\phi}_{9}\phi^{2}_{03}.\\
2\mu^{2}_{1}&=-2\gamma\left(v^{2}_{1}+v^{2}_{2}\right)-3\lambda_{4}\frac{v_{3}}{v_{1}}\left(v^{2}_{1}-v^{2}_{2}\right)-\lambda v^{2}_{3}+\lambda^{H\phi}_{1}\left(\phi^{2}_{01}+\phi^{2}_{02}\right)+\lambda^{H\phi}_{I}\frac{v_{3}}{2v_{2}}\left(\phi^{2}_{01}-\phi^{2
}_{02}\right)+\lambda^{H\phi}_{II}\frac{v_{1}}{v_{2}}
\phi_{01}\phi_{03}+\lambda^{H\phi}_{9}\phi^{2}_{03}.\\
2\mu^{2}_{2}&=-\lambda_{4}\frac{v_{2}}{v_{3}}\left(3v^{2}_{1}-v^{2}_{2}\right)-\lambda\left(v^{2}_{1}+v^{2}_{2}\right)-2\lambda_{8}v^{2}_{3}+\lambda^{H\phi}_{I}\frac{v_{2}}{2v_{3}}\left(\phi^{2}_{01}-\phi^{2
}_{02}\right)+\lambda^{H\phi}_{8}\left(\phi^{2}_{01}+\phi^{2
}_{02}\right)+\lambda^{H\phi}_{III}\frac{v_{1}}{2v_{3}}\phi_{01}\phi_{03}+\lambda^{H\phi}_{14}\phi^{2}_{03}.\\
2\mu^{2}_{1BL}&=-2\gamma_{BL}\left(\phi^{2}_{01}+\phi^{2}_{02}\right)-
\lambda_{BL}\phi^{2}_{03}+\lambda^{H \phi}_{1}\left(v^{2}_{1}+v^{2}_{2}\right)+\lambda^{H \phi}_{I} v_{2}v_{3}+\lambda^{H \phi}_{II}\frac{\phi_{03}}{\phi_{01}}v_{1}v_{2}+\lambda^{H \phi}_{8}v^{2}_{3}+\lambda^{H \phi}_{III}\frac{\phi_{03}}{\phi_{01}}v_{1}v_{3}.\\
2\mu^{2}_{1BL}&=-2\gamma_{BL}\left(\phi^{2}_{01}+\phi^{2}_{02}\right)-
\lambda_{BL}\phi^{2}_{03}+\lambda^{H \phi}_{1}\left(v^{2}_{1}+v^{2}_{2}\right)-\lambda^{H \phi}_{I} v_{2}v_{3}+\lambda^{H \phi}_{8}v^{2}_{3}.\\
2\mu^{2}_{2BL}&=-\lambda_{BL}\left(\phi^{2}_{01}+\phi^{2}_{02}\right)-2\lambda^{\phi}_{8}\phi^{2}_{03}+\lambda^{H \phi}_{II}\frac{\phi_{01}}{\phi_{03}}v_{1}v_{2}+\lambda^{H \phi}_{9}\left(v^{2}_{1}+v^{2}_{2}\right)+\lambda^{H \phi}_{III}\frac{\phi_{01}}{2\phi_{03}}v_{1}v_{3}+\lambda^{H\phi}_{14}v^{2}_{3}.
\end{align}}
where
\begin{align}
\lambda&= \lambda_{5}+\lambda_{6}+2\lambda_{7},\qquad \gamma=\lambda_{1}+\lambda_{3};\\
\lambda_{BL}&= \lambda^{\phi}_{5}+\lambda^{\phi}_{6}+2\lambda^{\phi}_{7},\qquad \gamma_{BL}=\lambda^{\phi}_{1}+\lambda^{\phi}_{3};\\
\lambda^{H\phi}_{I}&=\lambda^{H\phi}_{4}+\lambda^{H\phi}_{6},\qquad \lambda^{H\phi}_{II}=\lambda^{H\phi}_{5}+\lambda^{H\phi}_{7},\qquad \lambda^{H\phi}_{III}=\lambda^{H\phi}_{10}+\lambda^{H\phi}_{11}+\lambda^{H\phi}_{12}+\lambda^{H\phi}_{14}.
\end{align}

It is not the purpose of this paper to analyze the scalar potential in
detail, but some things can be noted.  The potential of the three
Higgs $S3$ model ($S3-3H$) has been analyzed in some detail in
refs.~\cite{Das:2014fea, Barradas-Guevara:2014yoa}. In our case, the
breaking of the $U(1)_{B-L}$ symmetry at a scale larger than the
electroweak scale $\phi_{0i} >> v_i$, will give rise to a massive
$Z_{B-L}$ gauge boson. After electroweak symmetry breaking the
remaining degrees of freedom from the $U(1)_{B-L}$ part will mix with
the ones coming from the electroweak doublets $H_i$, which transform
under $\bf{S}_3$, and give rise to a number of neutral, charged and
pseudoscalar Higgs bosons, one of which will correspond to the SM one.
This will provide two scales in the model, and some of the scalars
will be naturally heavier than the others, but it is clear from the
potential that there will be also mixing among them. This rich scalar
structure will give rise to FCNC's, a detailed analysis of which,
together with the experimental Higgs bounds, will place constraints on
the available parameter space of the model.  In the limit where the
couplings of the $B-L$ part go to zero the $S3-3H$ model will be
recovered. In general, the phenomenology of the models will be
different, not only because of the extra heavy scalar sector and a
$Z_{B-L}$ boson, but also because there may also  be mixing of the
$B-L$ sector and the $S3-3H$ one.

To get an idea what could be possible scenarios for the scalar
particles in our model let us consider the limit where there is no
mixing between the $B-L$ part and the $S3-3H$ one.  This situation will
correspond to two separate sectors at very different scales
$\phi_{0i}>>v_{i}$. As already mentioned, in the limit where the couplings of the $B-L$ part go to
zero the $S3-3H$ model will be recovered. In the $S3-3H$ model, after
electroweak symmetry breaking, the scalar sector consists of three neutral
scalars, $h_0$, $H_{1,2}$ one of which is identified with the Higgs
boson of the SM (say $H_2$), four charged scalars $H_{1,2}^{\pm}$ and
two pseudoscalars $A_{1,2}$.  There are two scenarios possible.  In
one case the $\lambda_4$ coupling  is absent, which implies a continuous
symmetry of the potential $SO(2)$. Upon breaking of the
electroweak symmetry this gives rise to a massless
Goldstone boson $h_0$\cite{Beltran:2009zz,Das:2014fea}, i.e. one of the
three neutral scalars remains massless.  The other two scalars $H_{1,2}$ can be
parameterized in a similar way to the two Higgs doublet model (2HDM)
and a decoupling or alignment limit defined.  This decoupling limit
refers to the fact that only one of the two scalars will be coupled to
the gauge bosons, and is identified with the SM one, the other scalar
is orthogonal to it and has no couplings with the gauge bosons.  But,
in contrast to the 2HDM, this decoupling limit does not imply that the
decoupled scalar is necessarily heavier than the other one.  On the
other hand, if the $\lambda_4$ term is present the continuous $SO(2)$
symmetry is not there, instead upon electroweak symmetry breaking
there is a residual $Z_2$ symmetry left from the breaking of $S_3$.
 Now the three neutral scalars acquire mass, but
one of them is not coupled to the gauge bosons, due to the $Z_2$
symmetry, and in the other two the decoupling limit described above
applies \cite{Das:2014fea}.  Although the three neutral scalars can
have masses in the same energy range, a study from a model with $\bf{S}_3$
symmetry and 4 Higgs doublets, where the fourth one is inert and the
couplings of the other three are like in $S3-3H$, shows that upon
certain considerations it is possible to satisfy the Higgs
bounds and have regions in parameter space that are compatible with
the latest experimental results \cite{Espinoza:2018itz}.

Examination of the $B-L$ part with no mixing terms shows that it
resembles the situation of the $S3-3H$ model with the $\lambda_4$ term
set to zero, that is, there exists an $SO(2)$ symmetry in this sector
too.  In this case, after the $\phi's$ acquire vev's, besides the
massive gauge boson $Z_{B-L}$, there will be three neutral scalars, one of
them massless, and two pseudoscalars.  The massive states will be
heavier than in the $S3-3H$ part, since we have assumed
$\phi_i>>v_{i}$, giving two disconnected scalar sectors and one
candidate to the SM Higgs boson in the decoupling limit described
above, plus the massless scalar.  In this case, since the
$\lambda_4^{\phi}$ coupling is forbidden by the $Z_2^e$ symmetry, 
the only way to avoid the Goldstone boson is to break softly this symmetry.

Upon considering the mixing of the $B-L$ part and the $S3-3H$ one, the
$SO(2)$ or $Z_2$ symmetries of the potential will not
be present, they will be broken by the mixing terms.  In general, all
the scalars will acquire masses.  Since the mixing terms have to be
very small to avoid the experimental bounds, it will still be possible
to define a decoupling limit in the sense described above, where one
of the neutral scalars of the $S3-3H$ part can be identified with the SM
one, although the expressions will be more complicated due to the
mixing terms.  The viability of this decoupling limit will impose constraints on the possible values of these
mixed couplings.

Moving to the fermionic sector, 
the Yukawa Lagrangian in the standard basis is 
\begin{align}
-\mathcal{L}_{Y}=\bar{q}_{i L} \left({\bf M}_{q} \right)_{ij}q_{j R}+\bar{\ell}_{i L} \left( {\bf M}_{\ell}\right)_{ij}\ell_{j R}
+\dfrac{1}{2}\bar{\nu}_{i L}\left({\bf M}_{\nu}\right)_{ij}\nu^{c}_{j L }+\dfrac{1}{2}\bar{N}^{c}_{i}\left({\bf M}_{R}\right)_{ij}N_{j }+h.c.\label{eq7}
\end{align}
where the type I see-saw mechanism has been realized, ${\bf
  M}_{\nu}=-{\bf M}_{D} {\bf M}^{-1}_{R} {\bf M}^{T}_{D}$. From
Eq.(\ref{eq2}), the mass matrices have the following form
\begin{align}
{\bf M}_{q}=\begin{pmatrix}
a_{q}+b^{\prime}_{q} & b_{q} & c_{q} \\ 
b_{q} & a_{q}-b^{\prime}_{q} & c^{\prime}_{q} \\ 
f_{q} & f^{\prime}_{q} & g_{q}
\end{pmatrix},\, {\bf M}_{\ell}=\begin{pmatrix}
a_{\ell} & 0 & 0 \\ 
0 & b_{\ell}+c_{\ell} & 0 \\ 
0 & 0 & b_{\ell}-c_{\ell}
\end{pmatrix}, \, {\bf M}_{R}=\begin{pmatrix}
a_{R} & b_{R} & b^{\prime}_{R} \\ 
b_{R} & c_{R} & 0 \\ 
b^{\prime}_{R} & 0 & c_{R}
\end{pmatrix},\label{eq8} 
\end{align}
where the $q= u, d$ and $\ell=e, D$. Explicitly, the matrix elements
for the quarks and lepton sectors are given as
\begin{align}
&a_{q}=y^{q}_{2}\langle H_{3}\rangle,\quad b^{\prime}_{q}=y^{q}_{1} \langle H_{2}\rangle, \quad b_{q}=y^{q}_{1} \langle H_{1}\rangle,\quad c_{q}=y^{q}_{3} \langle H_{1}\rangle,\quad
c^{\prime}_{q}=y^{q}_{3} \langle H_{2}\rangle,\quad f_{q}=y^{q}_{4} \langle H_{1}\rangle;\nn\\&f^{\prime}_{q}= y^{q}_{4} \langle H_{2}\rangle 
,\quad g_{q}=y^{q}_{5} \langle H_{3}\rangle,\quad
a_{\ell}=y^{\ell}_{1}\langle H_{3}\rangle,\quad b_{\ell}=y^{\ell}_{3}\langle H_{3}\rangle,\quad c_{\ell}=y^{\ell}_{2}\langle H_{2}\rangle,\quad
a_{R}=y^{N}_{1} \langle \phi_{3}\rangle;\nn\\&b_{R}=y^{N}_{2} \langle \phi_{1}\rangle,\quad b^{\prime}_{R}=y^{N}_{2} \langle \phi_{2}\rangle,\quad c_{R}=y^{N}_{3}\langle \phi_{3}\rangle .
\label{eq9}
\end{align}
In here, it is convenient to remark the number of Yukawa couplings that appear
in the flavored B-L model is reduced to half in comparison to the flavored LRSM scenario
\cite{Gomez-Izquierdo:2017rxi}.

\section{Masses and Mixings}

\subsection{Quark Sector: NNI Textures}
The quark mass matrix, ${\bf M}_{q}$, has already been obtained by means of the ${\bf S}_{3}$ flavor symmetry \cite{Kubo:2003iw, Kubo:2005sr, Mondragon:2006hi, Felix:2006pn,
Mondragon:2007af, Mondragon:2007nk,
Mondragon:2007jx,
Canales:2012dr, GonzalezCanales:2012za, Canales:2013ura,
Canales:2013cga}. However, it is important to point out, as it was shown in 
\cite{Canales:2013cga}, that this mass matrix possesses 
implicitly a
kind of NNI textures,\footnote{In \cite{Canales:2013cga}, the authors did not analyze completely this case neither diagonalize the mass matrix. They focused in kind of mass matrix with two zeroes like this 
\begin{equation}
{\bf M}=\divideontimes{\bf 1}+\begin{pmatrix}
0 & \star & 0 \\ 
\star^{\ast} & \star-\divideontimes & \star \\ 
0 & \star & \star-\divideontimes
\end{pmatrix}.\nonumber
\end{equation}}
but with one more free parameter than the 
canonical NNI ones \cite{Branco:1988iq, Branco:1994jx, Harayama:1996am, Harayama:1996jr} that only contain four. 
This is relevant since it shows that NNI textures
are hidden in the ${\bf S}_{3}$ flavor symmetry
\cite{Canales:2013cga}, so it may not be 
necessary to use larger discrete groups, for example the ${\bf Q}_{6}$
symmetry \cite{Babu:2004tn, Kajiyama:2005rk, Kajiyama:2007pr,
  Kifune:2007fj, Babu:2009nn, Kawashima:2009jv, Babu:2011mv}, to
understand the mixing by hierarchical mass matrices, although extending the symmetry group may be necessary in other contexts.
 
Having emphasized the above fact, we obtain
  simultaneously the NNI textures and the broken
  $\mu \leftrightarrow \tau$ symmetry, in the quark and lepton sector
  respectively, within an ${\bf S}_{3}$ flavored B-L gauge
  model. Although the NNI textures and the $\mu \leftrightarrow \tau$
  symmetry have been studied quite widely in the literature, both have
  not been explored in the present theoretical framework.

Let us comment on how to get the NNI textures (for
  more details on other textures see cite
  \cite{Canales:2013cga}). Taking the quark mass matrix,
  ${\bf M}_{q}$, that is diagonalized by the unitary matrices
  ${\bf U}_{q(R, L)}$ such that
  $\hat{\bf M}_{q}=\textrm{diag.}\left(m_{q_{1}}, m_{q_{2}},
    m_{q_{3}}\right)={\bf U}^{\dagger}_{q L}{\bf M}_{q}{\bf U}_{q
    R}$. Now, we apply the rotation ${\bf U}_{\theta}$
  (${\bf U}_{q(R, L)}={\bf U}_{\theta} {\bf u}_{q (R, L)}$) to
  ${\bf M}_{q}$ to obtain 
\begin{align}\label{NNI2}
{\bf m}_{q}={\bf U}^{T}_{\theta}{\bf M}_{q} {\bf U}_{\theta}=
\begin{pmatrix}
a_{q} & \frac{2}{\sqrt{3}}b^{\prime}_{q} & 0 \\ 
\frac{2}{\sqrt{3}}b^{\prime}_{q} & a_{q} & \frac{2}{\sqrt{3}}c^{\prime}_{q} \\ 
0 & \frac{2}{\sqrt{3}}f^{\prime}_{q} & g_{q}
\end{pmatrix},\,\, {\bf U}_{\theta}=\begin{pmatrix}
\cos{\theta} & \sin{\theta} & 0 \\ 
-\sin{\theta} & \cos{\theta} & 0 \\ 
0 & 0 & 1
\end{pmatrix},
\end{align}
with the following conditions
\begin{equation}
\tan{\theta}=\frac{c_{q}}{c^{\prime}_{q}}=\frac{f_{q}}{f^{\prime}_{q}}=\frac{\langle H_{1}\rangle}{\langle H_{2}\rangle}\qquad \textrm{and}\qquad \tan{2\theta}=\frac{b^{\prime}_{q}}{b_{q}}=\frac{\langle H_{2}\rangle}{\langle H_{1}\rangle}~,
\end{equation}
which give us the relation $\langle H_{2}\rangle=\pm\sqrt{3} \langle H_{1}\rangle$ \footnote{There is another way to get the NNI textures in ${\bf M}_{q}$ (see Eq.(\ref{eq8})). We know that $\hat{\bf M}_{q}=\textrm{diag.}\left(m_{q_{1}},
m_{q_{2}}, m_{q_{3}}\right)={\bf U}^{\dagger}_{q L}{\bf M}_{q}{\bf U}_{q R}$, if we assume that $\langle H_{2}\rangle=0$, we apply ${\bf U}_{12}$ (${\bf U}_{q(R, L)}={\bf U}_{12}{\bf u}_{q(R, L)})$ to the resultant mass matrix to end of having
\begin{equation}
{\bf m}_{q}={\bf U}^{T}_{12}{\bf M}_{q}{\bf U}_{12}
=\begin{pmatrix}
a_{q} & b_{q} & 0 \\ 
b_{q} & a_{q} & c_{q} \\ 
0 & f_{q} & g_{q}
\end{pmatrix},\qquad {\bf U}_{12}=\begin{pmatrix}
0 & 1 & 0 \\ 
1 & 0 & 0 \\ 
0 & 0 & 1 
\end{pmatrix},\nonumber 
\end{equation}
where ${\bf m}_{q}$ can be written as the Eq. (\ref{NNI3}). However,
the assumption $\langle H_{2} \rangle=0$ would imply exact
$\mu \leftrightarrow \tau$ symmetry in the charged leptons, which 
means, $m_{\mu}=m_{\tau}$.}, then $\theta=\pi/6$ as was shown in
\cite{Canales:2013cga}. Notice that ${\bf m}_{q}$ can be written as
\begin{equation}\label{NNI3}
{\bf m}_{q}=a_{q}{\bf 1}+\overbrace{\begin{pmatrix}
0 & \frac{2}{\sqrt{3}}b^{\prime}_{q} & 0 \\ 
\frac{2}{\sqrt{3}}b^{\prime}_{q} & 0 & \frac{2}{\sqrt{3}}c^{\prime}_{q} \\ 
0 & \frac{2}{\sqrt{3}}f^{\prime}_{q} & g_{q}-a_{q}
\end{pmatrix}}^{{\bf m^{\prime}_{q}}}.
\end{equation}
If ${\bf m}_{q}$ was a hermitian matrix
($f^{\prime}_{q}=c^{\prime \ast}_{q}$), this would imply that
${\bf u}^{\dagger}_{q L}{\bf u}_{q R}={\bf 1}$ and
${\bf m^{\prime}_{q}}$ would be like the Fritzsch textures so that to
diagonalize ${\bf m}_{q}$ is equivalent to do so in
${\bf m}^{\prime}_{q}$, this means,
$\hat{\bf M}^{\prime}_{q}=\textrm{diag.}\left(m_{q_{1}}-a_{q},
  m_{q_{2}}-a_{q}, m_{q_{3}}-a_{q}\right)={\bf u}^{\dagger}_{q L}{\bf
  m}^{\prime}_{q}{\bf u}_{q R}$. However, in the present framework,
${\bf m}_{q}$ is not hermitian and $a_{q}\neq 0$, in general, so an
exact diagonalization of ${\bf m}_{q}$ might produce a different
result from the one expected if $a_{q}=0$ (in this benchmark the
NNI textures appear). Along with this, if $a_{q}$ was considered as a
perturbation to ${\bf m}^{\prime}_{q}$, one would expect a modified
NNI texture. In here, for simplicity, in order to not include extra
discrete symmetries to prohibit the second term in the Yukawa mass
term (see Eq.(\ref{eq2})), which gives rise to $a_{q}$, let us adopt
the benchmark where $a_{q}=0$ which means that $y^{q}_{2}=0$. In this
way, the NNI textures appear in the quark mass matrix so this hierarchical
matrices fit very well the CKM matrix.

In this framework, we find out the ${\bf u}_{f R}$ and ${\bf u}_{f L}$
unitary matrices that diagonalize ${\bf m}_{q}$. Then, we must build the
bilineal forms: ${\bf \hat{M}}_{q} {\bf \hat{M}}^{\dagger}_{q}={\bf
  u}^{\dagger}_{q L} {\bf m}_{q} {\bf m }^{\dagger}_{q} {\bf u}_{q L}$
and ${\bf \hat{M}}^{\dagger}_{q} {\bf \hat{M}}_{q}={\bf
  u}^{\dagger}_{q R} {\bf m}^{\dagger}_{q} {\bf m }_{q} {\bf u}_{q
  R}$, however, in this work we will only need to  obtain the
${\bf u}_{q L}$ left-handed matrix which takes place in the CKM
matrix. This is given by ${\bf u}_{q L}={\bf Q}_{q L}{\bf O}_{q L}$
where the former matrix contains the CP-violating phases, $ {\bf
  Q}_{q} = \textrm{diag} \left( 1, \exp i\eta_{q_{2}}, \exp
  i\eta_{q_{3}} \right)$, that comes from ${\bf m}_{q} {\bf m
}^{\dagger}_{q}$. ${\bf O}_{q L}$ is a real orthogonal matrix and it
is parametrized as
\begin{align} \label{ortho}
{\bf O}_{q L}= 
\begin{pmatrix}
-\sqrt{\dfrac{\tilde{m}_{q_{2}} (\rho^{q}_{-}-R^{q}) K^{q}_{+}}{4 y_{q} \delta^{q}_{1} \kappa^{q}_{1} }} & -\sqrt{\dfrac{\tilde{m}_{q_{1}} (\sigma^{q}_{+}-R^{q}) K^{f}_{+}}{4 y_{q} \delta^{q}_{2} \kappa^{q}_{2} }}  & \sqrt{\dfrac{\tilde{m}_{q_{1}} \tilde{m}_{q_{2}} (\sigma^{q}_{-}+R^{q}) K^{q}_{+}}{4 y_{q} \delta^{q}_{3} \kappa^{q}_{3} }} \\ 
-\sqrt{\dfrac{\tilde{m}_{q_{1}} \kappa^{q}_{1} K^{q}_{-}}{\delta^{q}_{1}(\rho^{q}_{-}-R^{q}) }} & \sqrt{\dfrac{\tilde{m}_{q_{2}} \kappa^{q}_{2} K^{q}_{-}}{\delta^{q}_{2}(\sigma^{q}_{+}-R^{q}) }} & \sqrt{\dfrac{\kappa^{q}_{3} K^{q}_{-}}{\delta^{q}_{3}(\sigma^{q}_{-}+R^{q}) }} \\ 
\sqrt{\dfrac{\tilde{m}_{q_{1}} \kappa^{q}_{1}(\rho^{q}_{-}-R^{q})}{2 y_{q}\delta^{q}_{1}}} & -\sqrt{\dfrac{\tilde{m}_{q_{2}} \kappa^{q}_{2}(\sigma^{q}_{+}-R^{q})}{2 y_{q}\delta^{q}_{2}}}  & \sqrt{\dfrac{\kappa^{q}_{3}(\sigma^{q}_{-}+R^{q})}{2 y_{q}\delta^{q}_{3}}}
\end{pmatrix}
\end{align}
with
\begin{align} \label{defortho}
\rho^{q}_{\pm}&\equiv 1+\tilde{m}^{2}_{q_{2}}\pm\tilde{m}^{2}_{q_{1}}-y^{2}_{q},\quad \sigma^{q}_{\pm}\equiv 1-\tilde{m}^{2}_{q_{2}}\pm(\tilde{m}^{2}_{q_{1}}-y^{2}_{q}),\quad
\delta^{q}_{(1, 2)}\equiv (1-\tilde{m}^{2}_{q_{(1, 2)}})(\tilde{m}^{2}_{q_{2}}-\tilde{m}^{2}_{q_{1}});\nn\\
\delta^{q}_{3}&\equiv (1-\tilde{m}^{2}_{q_{1}})(1-\tilde{m}^{2}_{q_{2}}),\quad \kappa^{q}_{1} \equiv  \tilde{m}_{q_{2}}-\tilde{m}_{q_{1}}y_{q},\quad \kappa^{q}_{2}\equiv \tilde{m}_{q_{2}}y_{q}-\tilde{m}_{q_{1}},\quad \kappa^{q}_{3}\equiv y_{q}-\tilde{m}_{q_{1}}\tilde{m}_{q_{2}};\nn\\
R^{q}&\equiv \sqrt{\rho^{q 2}_{+}-4(\tilde{m}^{2}_{q_{2}}+\tilde{m}^{2}_{q_{1}}+\tilde{m}^{2}_{q_{2}}\tilde{m}^{2}_{q_{1}}-2\tilde{m}_{q_{1}}\tilde{m}_{q_{2}}y_{q})},\quad	K^{q}_{\pm} \equiv  y_{q}(\rho^{q}_{+}\pm R^{q})-2\tilde{m}_{q_{1}}\tilde{m}_{q_{2}}.
\end{align}
In the above expressions, all the parameters have been normalized by
the heaviest physical quark mass, $m_{q_{3}}$. Along with this, from
the above parametrization, $y_{q}\equiv \vt g_{q}\vt/m_{q_{3}}$, is
the only dimensionless free parameter that cannot be fixed in terms of
the physical masses, but is is constrained by, 
$1>y_{q}>\tilde{m}_{q_{2}}>\tilde{m}_{q_{1}}$. Therefore, the
left-handed mixing matrix that takes places in the CKM matrix is given
by ${\bf U}_{q L}= {\bf U}_{\theta}{\bf Q}_{q}{\bf O}_{q L}$ where
$q=u, d$. Finally, the CKM mixing matrix is written as
\begin{equation}
{\bf V}_{PMNS}={\bf O}^{T}_{u L}{\bf P}_{q} {\bf O}_{d L}, \quad {\bf P}_{q}={\bf Q}^{\dagger}_{u}{\bf Q}_{d}=\textrm{diag.}\left(1, e^{i\eta_{q_{1}}}, e^{i\eta_{q_{2}}} \right).
\end{equation}
This CKM mixing matrix has four free parameters, namely $y_{u}$,
$y_{d}$, and two phases $\eta_{q_{1}}$ and $\eta_{q_{2}}$ which could
be obtained numerically; in this work, the physical quark masses (at
$m_{Z}$ scale) will be taken (just central values) as inputs:
$m_{u}=1.45~MeV$, $m_{c}=635~MeV$, $m_{t}=172.1~GeV$ and
$m_{d}=2.9~MeV$, $m_{s}=57.7~MeV$, $m_{b}=2.82~GeV$
\cite{Bora:2012tx}. In the following, a naive $\chi^{2}$ analysis will
be performed to tune the free parameters. Then, we define
\begin{equation}
\chi^{2}\left(y_{u},y_{d}, \eta_{q_{1}}, \eta_{q_{2}}\right)=\frac{\left(\left| V^{th}_{ud}\right|-V^{ex}_{ud}\right)^{2}}{\sigma^{2}_{ud}}+\frac{\left(\left|V^{th}_{us}\right|-V^{ex}_{us}\right)^{2}}{\sigma^{2}_{us}}+\frac{\left(\left|V^{th}_{ub}\right|-V^{ex}_{ub}\right)^{2}}{\sigma^{2}_{ub}}+\frac{\left(\left|J^{th}\right|-J^{ex}\right)^{2}}{\sigma^{2}_{J}}.
\end{equation}
where the experimental values are given as \cite{Patrignani:2016xqp}
\begin{equation}
V^{ex}_{ud}=0.97434^{+0.00011}_{-0.00012},\quad V^{ex}_{us}=0.22506\pm 0.00050,\quad V^{ex}_{ub}=0.00357\pm 0.00015,\nonumber
\end{equation}
and
\begin{equation}
J^{th}=Im\left[V^{th}_{us}~V^{th}_{cb}~ V^{\ast th}_{cs}~ V^{\ast th}_{ub} \right],\quad J^{ex}=3.04^{+0.21}_{-0.20}\times 10^{-5}.\nonumber
\end{equation}
Then, we obtain the following values for the free parameters that fit the mixing values up to $2\sigma$
\begin{equation}
y_{u}=0.996068,\quad y_{d}=0.922299,\quad \eta_{q_{1}}=4.48161,\quad \eta_{q_{2}}=3.64887,
\end{equation}
with these values, one obtains
\begin{equation}
\left|V^{th}_{CKM}\right|=\begin{pmatrix}
0.97433 & 0.22505 & 0.00356 \\ 
0.22490 & 0.97359 & 0.03926 \\ 
0.00901 & 0.03831 & 0.99922
\end{pmatrix},\qquad J^{th}=3.04008\times 10^{-5}.
\end{equation}
As can be seen, these values are in good agreement with the experimental data, this is not a surprise since the NNI textures work quite well in the quark sector.

\subsection{Lepton Sector: Broken $\mu \leftrightarrow \tau$ Symmetry}

As we already mentioned, the lepton mass matrices have been already
diagonalized in the framework of the LRSM \cite{Gomez-Izquierdo:2017rxi},
where a systematic study was realized on the mixing angles. So, we
will just mention the relevant points and comment on the results.

The ${\bf M}_{e}$ mass matrix is complex and diagonal, then one can 
identify straightforwardly the physical masses;  since the ${\bf
M}_{e}$ mass
matrix is diagonalized by ${\bf U}_{e L}={\bf S}_{23}{\bf P}_{e}$ and
${\bf U}_{e R}={\bf S}_{23}{\bf P}^{\dg}_{e}$, this is, ${\hat{\bf
    M}_{e}}=\textrm{Diag.}(\vt m_{e}\vt, \vt m_{\mu}\vt,\vt
m_{\tau}\vt)={\bf U}^{\dg}_{e L}{\bf M}_{e}{\bf U}_{e R} ={\bf
  P}^{\dg}{\bf m}_{e}{\bf P}^{\dg}_{e}$ with ${\bf m}_{e}={\bf
  S}^{T}_{23}{\bf M}_{e}{\bf S}_{23}$. After factorizing the phases,
we have ${\bf m}_{e}={\bf P}_{e}{\bf \bar{m}_{e}}{\bf P}_{e}$~ where
 \begin{align}
 	{\bf m}_{e}=\textrm{Diag.}(m_{e}, m_{\mu}, m_{\tau}),\quad {\bf S}_{23}=\begin{pmatrix}
 	1 & 0 & 0 \\ 
 	0 & 0 & 1 \\ 
 	0 & 1 & 0
 	\end{pmatrix},\quad {\bf P}_{e}=\textrm{diag.}(e^{i\eta_{e}/2}, e^{i\eta_{\mu}/2}, e^{i\eta_{\tau}/2}) \label{eue}
 	\end{align}	
As a result, one obtains $\vt m_{e}\vt=\vt a_{e}\vt$, $\vt m_{\mu}\vt=\vt b_{e}-c_{e}\vt$ and $\vt m_{\tau}\vt=\vt b_{e}+c_{e}\vt$.

On the other hand, the ${\bf M}_{\nu}={\bf M}_{D}{\bf M}^{-1}_{R}{\bf M}^{T}_{D}$ effective neutrino mass matrix
is given by 
{\footnotesize
\begin{align}
{\bf M}_{\nu}=\begin{pmatrix}
\mc{X}a^{2}_{D}& -a_{D}\mc{Y}(b_{D}+c_{D})  & -a_{D}\mc{Y}(b_{D}-c_{D}) \\ 
-a_{D}\mc{Y}(b_{D}+c_{D})& \mc{W}(b_{D}+c_{D})^{2}  & \mc{Z}(b^{2}_{D}-c^{2}_{D})  \\ 
-a_{D}\mc{Y}(b_{D}-c_{D})& \mc{Z}(b^{2}_{D}-c^{2}_{D})  & \mc{W}(b_{D}-c_{D})^{2}
\end{pmatrix},\quad {\bf M}^{-1}_{R}\equiv\begin{pmatrix}
\mc{X}& -\mc{Y} & -\mc{Y} \\ 
-\mc{Y} & \mc{W} & \mc{Z} \\ 
-\mc{Y} & \mc{Z} & \mc{W} \label{efm}
\end{pmatrix} ,
\end{align}}
where the Dirac ($\ell=D$) and right-handed neutrino mass matrices are given in Eq. (\ref{eq8}). In the latter mass matrix, we have assumed the following vacuum alignment $\langle \phi_{1}\rangle=\langle \phi_{2}\rangle$. Now as hypothesis, we will assume that $b_{D}$ is larger than $c_{D}$, in this way the effective mass matrix can be written as
\begin{align}
{\bf M}_{\nu}\equiv\begin{pmatrix}
m^{0}_{ee}& -m^{0}_{e\mu}(1+\epsilon) & -m^{0}_{e\mu}(1-\epsilon) \\ 
-m^{0}_{e\mu}(1+\epsilon)& m^{0}_{\mu\mu}(1+\epsilon)^{2} & m^{0}_{\mu\tau}(1-\epsilon^{2}) \\ 
-m^{0}_{e\mu}(1-\epsilon)& m^{0}_{\mu\tau}(1-\epsilon^{2})  & m^{0}_{\mu \mu}(1-\epsilon)^{2}\label{efm2}
\end{pmatrix} ,
\end{align}
where $m^{0}_{ee}\equiv \mc{X}a^{2}_{D}$, $m^{0}_{e\mu}
\equiv\mc{Y}a_{D}b_{D}$, $m^{0}_{\mu \mu}\equiv\mc{W}b^{2}_{D}$ and
$m^{0}_{\mu\tau}\equiv\mc{Z}b^{2}_{D}$ are complex. In here,
$\epsilon\equiv c_{D}/b_{D}$ is a complex parameter which will be
considered as a perturbation to the effective mass matrix such that
$\vert \epsilon \vert \lll 1$. In order to break softly the $\mu
\leftrightarrow \tau$ symmetry, we demand that $\vert \epsilon
\vert\leq 0.3$ so we will neglect the $\epsilon^{2}$ quadratic terms
in the above matrix hereafter and a perturbative diagonalization will
be carried out.

In order to cancel the ${\bf S}_{23}$ contribution that comes from the
charged lepton sector, we proceed as follows. We
know that $\hat{\bf M}_{\nu}=\textrm{diag.}(m_{\nu_{1}}, m_{\nu_{2}},
m_{\nu_{3}})={\bf U}^{\dg}_{\nu}{\bf M}_{\nu}{\bf U}^{\ast}_{\nu}$,
then ${\bf U}_{\nu}={\bf S}_{23}{\bf \mc{U}_{\nu}}$ where the latter
mixing matrix will be obtained below. Then, $\hat{\bf M}_{\nu}={\bf
  \mc{U}^{\dg}_{\nu}}{\bf \mc{M}_{\nu}}{\bf \mc{U}^{\ast}_{\nu}}$ with
\begin{align}\label{efm3}
{\bf \mc{M}_{\nu}}\approx\begin{pmatrix}
	m^{0}_{e e} & -m^{0}_{e\mu} & -m^{0}_{e\mu} \\ 
	-m^{0}_{e\mu} & m^{0}_{\mu\mu} & m^{0}_{\mu\tau} \\ 
	-m^{0}_{e\mu} & m^{0}_{\mu\tau} & m^{0}_{\mu\mu}
	\end{pmatrix}+\begin{pmatrix}
	0 & m^{0}_{e \mu}~ \epsilon & -m^{0}_{e\mu} ~\epsilon \\ 
	-m^{0}_{e\mu}~ \epsilon & -2 m^{0}_{\mu\mu}~ \epsilon & 0 \\ 
	-m^{0}_{e\mu}~ \epsilon & 0 & 2 m^{0}_{\mu\mu}~ \epsilon
	\end{pmatrix}={\bf \mc{M}^{0}_{\nu}}+{\bf \mc{M}^{\epsilon}_{\nu}}.
\end{align}	
Notice that ${\bf \mc{M}^{0}_{\nu}}$ possesses the $\mu-\tau$ symmetry and this is diagonalized by
\begin{align}
{\bf \mc{U}}^{0}_{\nu}=\begin{pmatrix}
\cos{\theta}_{\nu}~e^{i\eta_{\nu}} & \sin{\theta}_{\nu}~e^{i\eta_{\nu}}  & 0 \\ 
-\frac{\sin{\theta}_{\nu}}{\sqrt{2}}& \frac{\cos{\theta}_{\nu}}{\sqrt{2}} & -\frac{1}{\sqrt{2}} \\ 
-\frac{\sin{\theta}_{\nu}}{\sqrt{2}}& \frac{\cos{\theta}_{\nu}}{\sqrt{2}} & \frac{1}{\sqrt{2}}
\end{pmatrix} \label{ubm} ,
\end{align}	
where the ${\bf \mc{M}^{0}_{\nu}}= {\bf \mc{U}}^{0}_{\nu} \hat{\bf M}^{0}_{\nu} {\bf \mc{U}}^{0 T}_{\nu} $ matrix elements are written as
\begin{align}
m^{0}_{e e}&=(m^{0}_{\nu_{1}}\cos^{2}{\theta}_{\nu}+m^{0}_{\nu_{2}}\sin^{2}{\theta}_{\nu})e^{2i\eta_{\nu}},\quad -m^{0}_{e\mu}=\frac{1}{\sqrt{2}}\cos{\theta}_{\nu}\sin{\theta}_{\nu}(m^{0}_{\nu_{2}}-m^{0}_{\nu_{1}})e^{i\eta_{\nu}};\nn\\ m^{0}_{\mu\mu}&=\frac{1}{2}(m^{0}_{\nu_{1}}\sin^{2}{\theta}_{\nu}+m^{0}_{\nu_{2}}\cos^{2}{\theta}_{\nu}+m^{0}_{\nu_{3}}),\quad
m^{0}_{\mu\tau}=\frac{1}{2}(m^{0}_{\nu_{1}}\sin^{2}{\theta}_{\nu}+m^{0}_{\nu_{2}}\cos^{2}{\theta}_{\nu}-m^{0}_{\nu_{3}}).\label{mne}
\end{align}	
Including the perturbation, ${\bf \mc{M}^{\epsilon}_{\nu}}$, applying ${\bf\mc{U}^{0}_{\nu}}$ one gets ${\bf \mc{M}}_{\nu}={\bf \mc{U}^{0\dg}_{\nu}}({\bf \mc{M}^{0}_{\nu}}+{\bf \mc{M}^{\epsilon}_{\nu}}){\bf \mc{U}^{0 \ast}_{\nu}}$. Explicitly
{\footnotesize
\begin{align}
{\bf \mc{M}}_{\nu}=\textrm{Diag.}(m^{0}_{\nu_{1}}, m^{0}_{\nu_{2}}, m^{0}_{\nu_{3}})+\begin{pmatrix}
0 & 0 &-\sin{\theta_{\nu}}(m^{0}_{\nu_{3}}+m^{0}_{\nu_{1}})~\epsilon \\ 
0 & 0 & \cos{\theta_{\nu}}(m^{0}_{\nu_{3}}+m^{0}_{\nu_{2}})~\epsilon \\ 
-\sin{\theta_{\nu}}(m^{0}_{\nu_{3}}+m^{0}_{\nu_{1}})~\epsilon& \cos{\theta_{\nu}}(m^{0}_{\nu_{3}}+m^{0}_{\nu_{2}})~\epsilon & 0\label{mnp}
\end{pmatrix} .
\end{align}	}
The contribution of second matrix to the mixing one is given by
\begin{align}
{\bf \mc{U}}^{\epsilon}_{\nu}\approx\begin{pmatrix}
N_{1}&  0 & -N_{3}\sin{\theta}~r_{1}~\epsilon \\ 
0 & N_{2} & N_{3}\cos{\theta_{\nu}}~r_{2}~\epsilon \\ 
N_{1}\sin{\theta_{\nu}}~r_{1}~\epsilon & -N_{2}\cos{\theta_{\nu}}~r_{2}~\epsilon & N_{3}
\end{pmatrix}, \label{ubpr}
\end{align}
where the $N_{1}$, $N_{2}$ and $N_{3}$ are the normalization factors which are given as
{\footnotesize
\begin{align}
N_{1}=\frac{1}{\sqrt{1+\sin^{2}{\theta_{\nu}}\vert r_{1}\epsilon \vert^{2}}} ,\quad N_{2}=\frac{1}{\sqrt{1+\cos^{2}{\theta_{\nu}}\vert r_{2}\epsilon \vert^{2}}},\quad N_{3}=\frac{1}{\sqrt{1+\sin^{2}{\theta_{\nu}}\vert r_{1}\epsilon \vert^{2}+\cos^{2}{\theta_{\nu}}\vert r_{2}\epsilon \vert^{2}}},\label{nofac}
\end{align}}
with $r_{(1, 2)}\equiv (m^{0}_{\nu_{3}}+m^{0}_{\nu_{(1, 2)}})/(m^{0}_{\nu_{3}}-m^{0}_{\nu_{(1, 2)}})$. Finally, the effective mass matrix given in Eq.(\ref{efm2}) is diagonalized approximately by ${\bf U}_{\nu}\approx {\bf S}_{23}{\bf \mc{U}^{0}_{\nu}{\bf \mc{U}^{\epsilon}_{\nu}}}$. Therefore, the theoretical PMNS mixing matrix is written as $V_{PMNS}={\bf U}^{\dg}_{e L}{\bf U}_{\nu}={\bf P}^{\dg}_{e}{\bf \mc{U}^{0}_{\nu}{\bf \mc{U}^{\epsilon}_{\nu}}}$. Explicitly,
\begin{align}
{\bf V}_{PMNS}={\bf P}^{\prime\dagger}_{e}\begin{pmatrix}
\cos{\theta_{\nu}}N_{1} & \sin{\theta_{\nu}}N_{2} & \sin{2\theta_{\nu}}\frac{N_{3}}{2}(r_{2}-r_{1})~\epsilon \\ 
-\frac{\sin{\theta_{\nu}}}{\sqrt{2}}N_{1}(1+r_{1}~\epsilon)& \frac{\cos{\theta_{\nu}}}{\sqrt{2}}N_{2}(1+r_{2}~\epsilon) & -\frac{N_{3}}{\sqrt{2}}\left[1-\epsilon~r_{3}\right] \\ 
-\frac{\sin{\theta_{\nu}}}{\sqrt{2}}N_{1}(1-r_{1}~\epsilon)& \frac{\cos{\theta_{\nu}}}{\sqrt{2}}N_{2}(1-r_{2}~\epsilon)  & \frac{N_{3}}{\sqrt{2}}\left[1+\epsilon~ r_{3}\right]
\end{pmatrix} ,\label{pmma}
\end{align}	
where the Dirac phase, $\eta_{\nu}$, has been factorized in the first entry of ${\bf P}^{\prime\dagger}_{e}$ and  $r_{3}\equiv r_{2}\cos^{2}{\theta_{\nu}}+r_{1}\sin^{2}{\theta_{\nu}}$. On the other hand,
comparing the magnitude of entries in the ${\bf V}_{PMNS}$ with the mixing matrix in the 
standard parametrization of the PMNS, we obtain the following expressions for 
the lepton mixing angles
\begin{align}
\sin^{2}{\theta}_{13}&=\vert {\bf V}_{13}\vert^{2} =\frac{\sin^{2}{2\theta_{\nu}}}{4}N^{2}_{3}\vert \epsilon \vert^{2}~\vert r_{2}-r_{1} \vert^{2};\nn\\
\sin^{2}{\theta}_{23}&=\dfrac{\vert {\bf V}_{23}\vert^{2}}{1-\vert {\bf V}_{13}\vert^{2}}=\dfrac{N^{2}_{3}}{2}\frac{\vert 1-\epsilon~r_{3} \vert ^{2}}{1- \sin^{2}{\theta_{13}}},\nn\\
\sin^{2}{\theta_{12}}&=\dfrac{\vert {\bf V}_{12}\vert^{2}}{1-\vert {\bf V}_{13}\vert^{2}}=
\dfrac{N^{2}_{2}\sin^{2}{\theta_{\nu}}}{1-\sin^{2}{\theta}_{13}}.\label{mixang}
\end{align} 

Notice that, in general, the reactor and atmospheric angles depend
strongly on the active neutrino masses and therefore on the Majorana
phases; also the reactor angle depends on the magnitude of the
parameter $\epsilon$ but the atmospheric one has a clear dependency on
the $\epsilon$ phase, which turns out to be relevant for reaching the allowed
value.

In particular, as it was shown in \cite{Gomez-Izquierdo:2017rxi}, in the
regime of a soft breaking of the $\mu \leftrightarrow \tau$ symmetry,
as one can see from Eq. (\ref{mixang}), $\theta_{12}\approx
\theta_{\nu}$, then this parameter was considered as an input to
determine the reactor and atmospheric angles. In these circumstances,
the normal hierarchy was ruled out by experimental data.
Along with this, the
most viable cases for inverted and degenerate hierarchy were those
where CP parities in the neutrino masses are ${\bf
\mc{M}^{0}_{\nu}}=\textrm{diag.}\left(m^{0}_{\nu_{1}},
m^{0}_{\nu_{2}}, m^{0}_{\nu_{3}} \right)=\textrm{diag.}\left(-
\left|m^{0}_{\nu_{1}}\right|,
\left|m^{0}_{\nu_{2}}\right|,-\left|m^{0}_{\nu_{3}}\right| \right)$
where
\begin{eqnarray}
\vert m^{0}_{\nu_{2}}\vert&=&\sqrt{\Delta m^{2}_{13}+\Delta m^{2}_{21}+\vert m^{0 }_{\nu_{3}}\vert^{2}},\qquad\vert m^{0}_{\nu_{1}}\vert=\sqrt{\Delta m^{2}_{13}+\vert m^{0}_{\nu_{3}}\vert^{2}},\qquad \textrm{Inverted Hierarchy}\nonumber\\
\vert m^{0}_{\nu_{3}} \vert&=& \sqrt{\Delta m^{2}_{31}+m^{2}_{0}},\qquad \vert m^{0}_{\nu_{2}} \vert=\sqrt{\Delta m^{2}_{21}+m^{2}_{0}}, \qquad \textrm{Degenerate Hierarchy}
\end{eqnarray}
with $m_{0}\gtrsim 0.1~eV$ as the common mass.  At the same time, for
the inverted (degenerate) hierarchy the associated phase of
$\epsilon=\left|\epsilon\right|e^{\alpha_{\epsilon}}$ has to be $0$
($\pi$) to reach the allowed values for the reactor and atmospheric
angles.
 
\begin{figure}[ht]\centering
\includegraphics[scale=0.44]{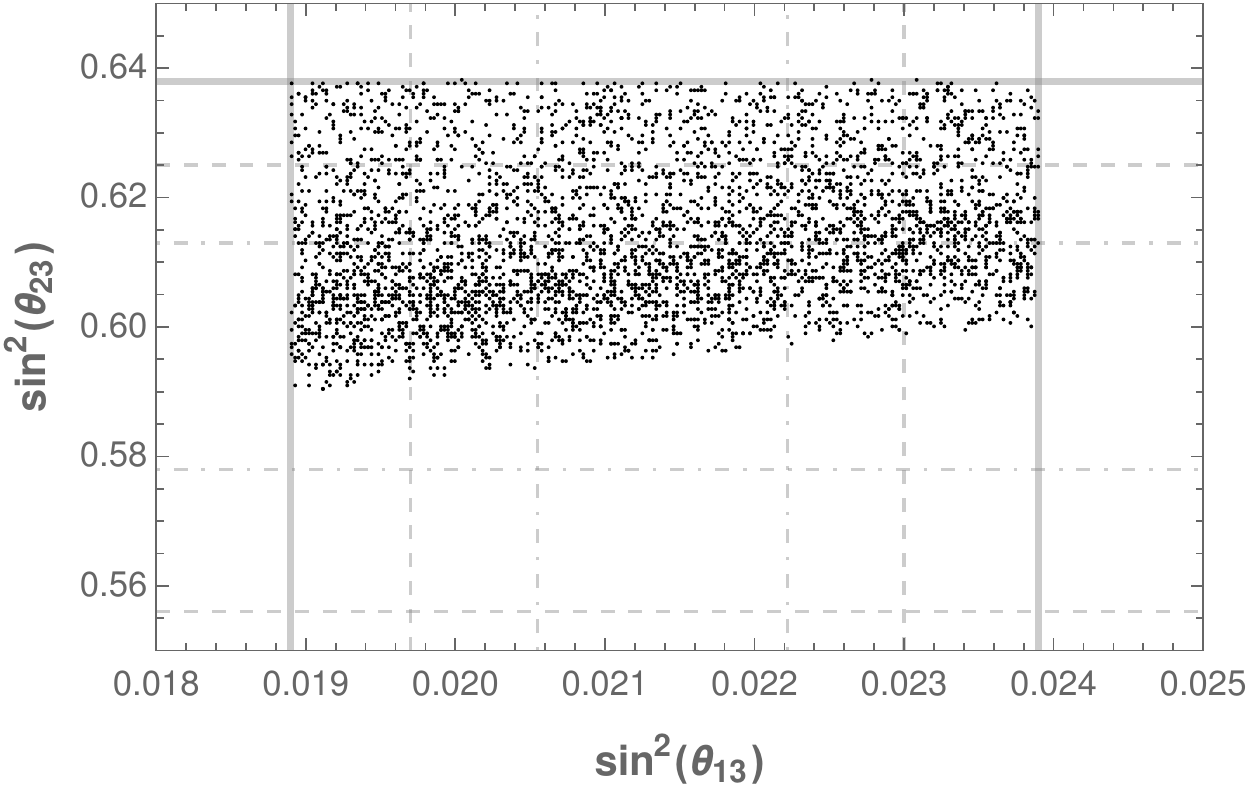}
\hspace{1mm}\includegraphics[scale=0.41]{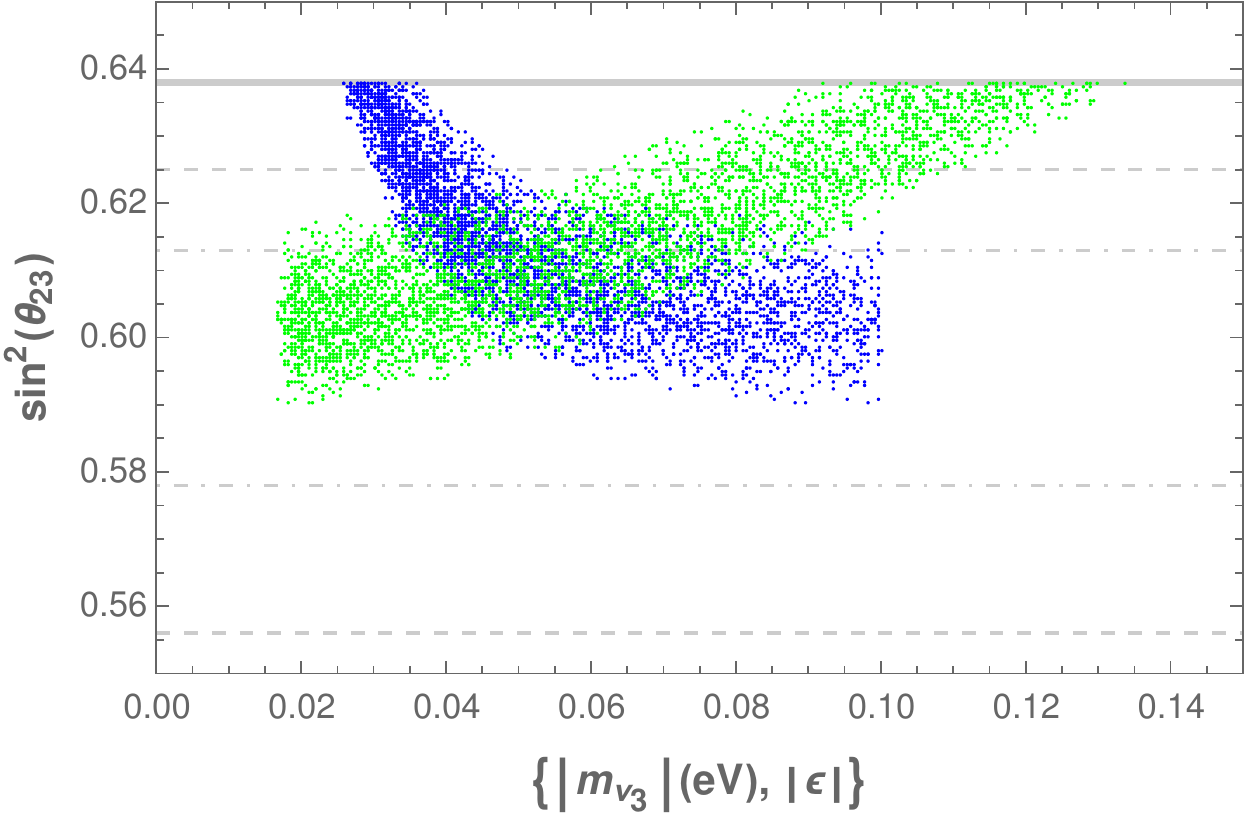}
\hspace{1mm}\includegraphics[scale=0.42]{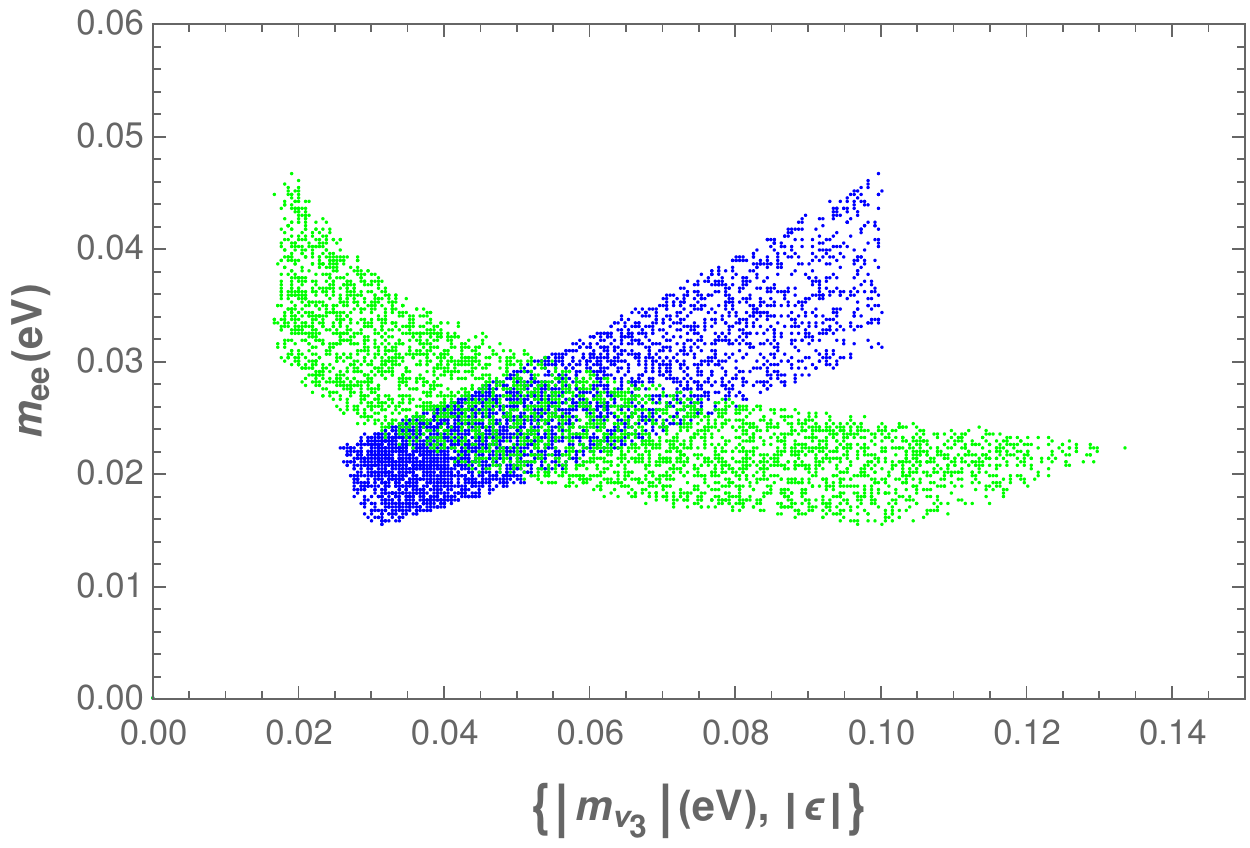}
\caption{From left to right:  the atmospheric angle versus (a) the reactor angle , (b) $\vert
m^0_{\nu_{3}}\vert$  and $\left|\epsilon
\right|$, and (c) $m_{ee}$ versus $\vert
m^0_{\nu_{3}}\vert$  and $\left|\epsilon \right|$. Inverted hierarchy: blue and green points stand for
$\left|m_{\nu_{3}}\right|$ and $\left|\epsilon\right|$,
respectively. The dotdashed, dashed and thick lines stand for $1~\sigma$, $2~\sigma$ and $3~\sigma$ of C. L.}\label{mee6}
\end{figure}

\begin{figure}[ht]\centering
\includegraphics[scale=0.42]{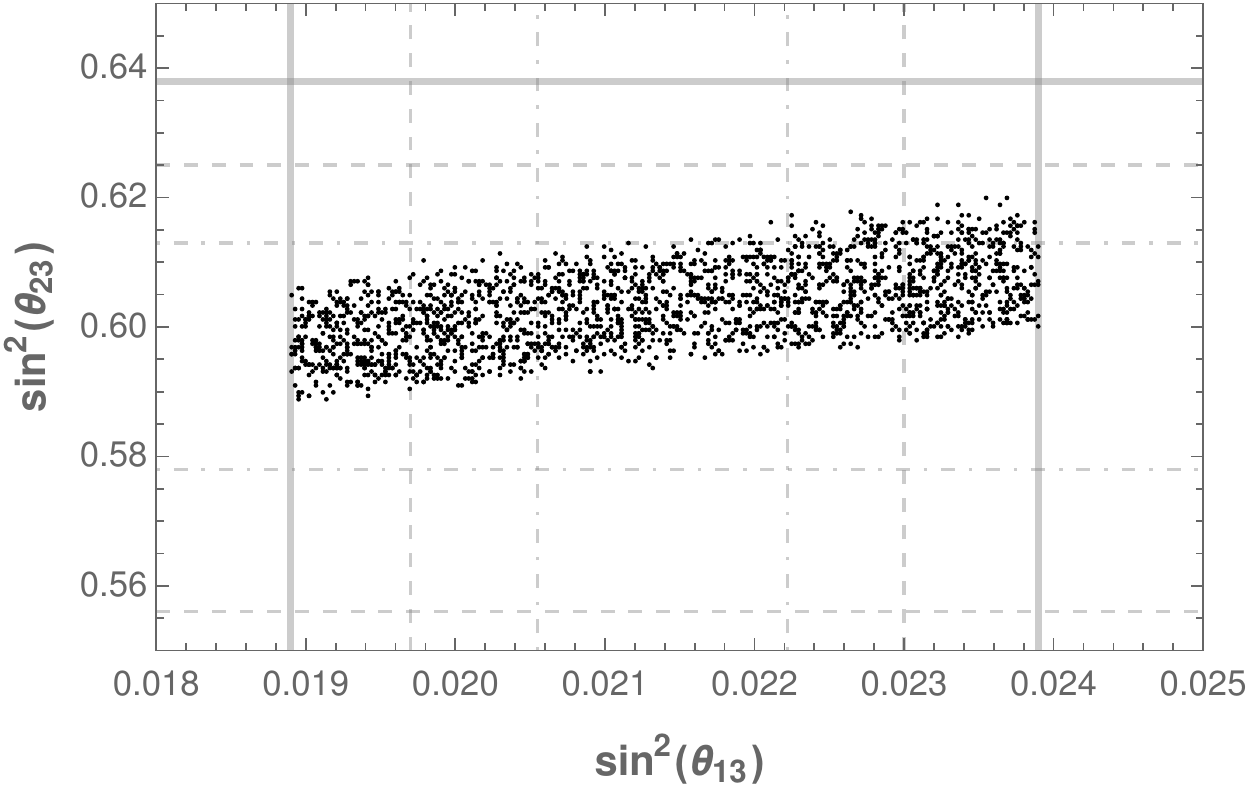}
\hspace{1mm}\includegraphics[scale=0.43]{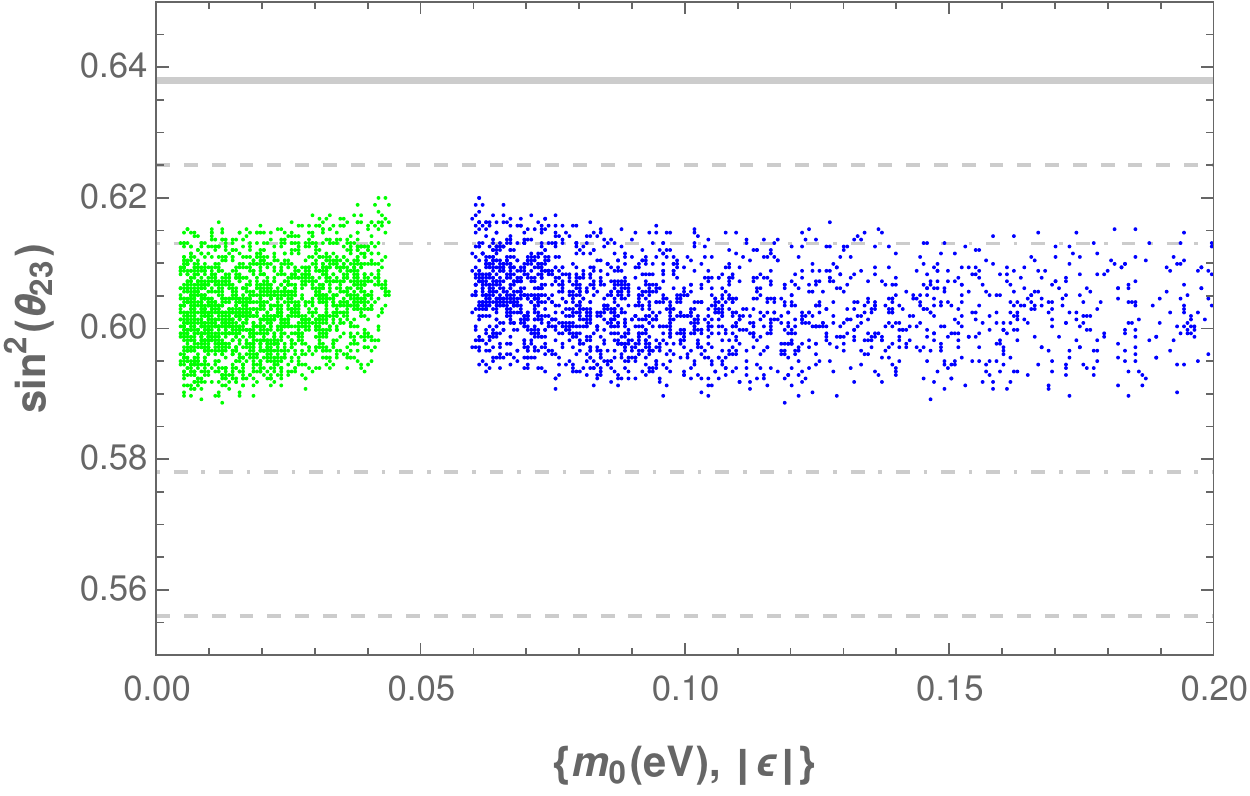}
\hspace{1mm}\includegraphics[scale=0.42]{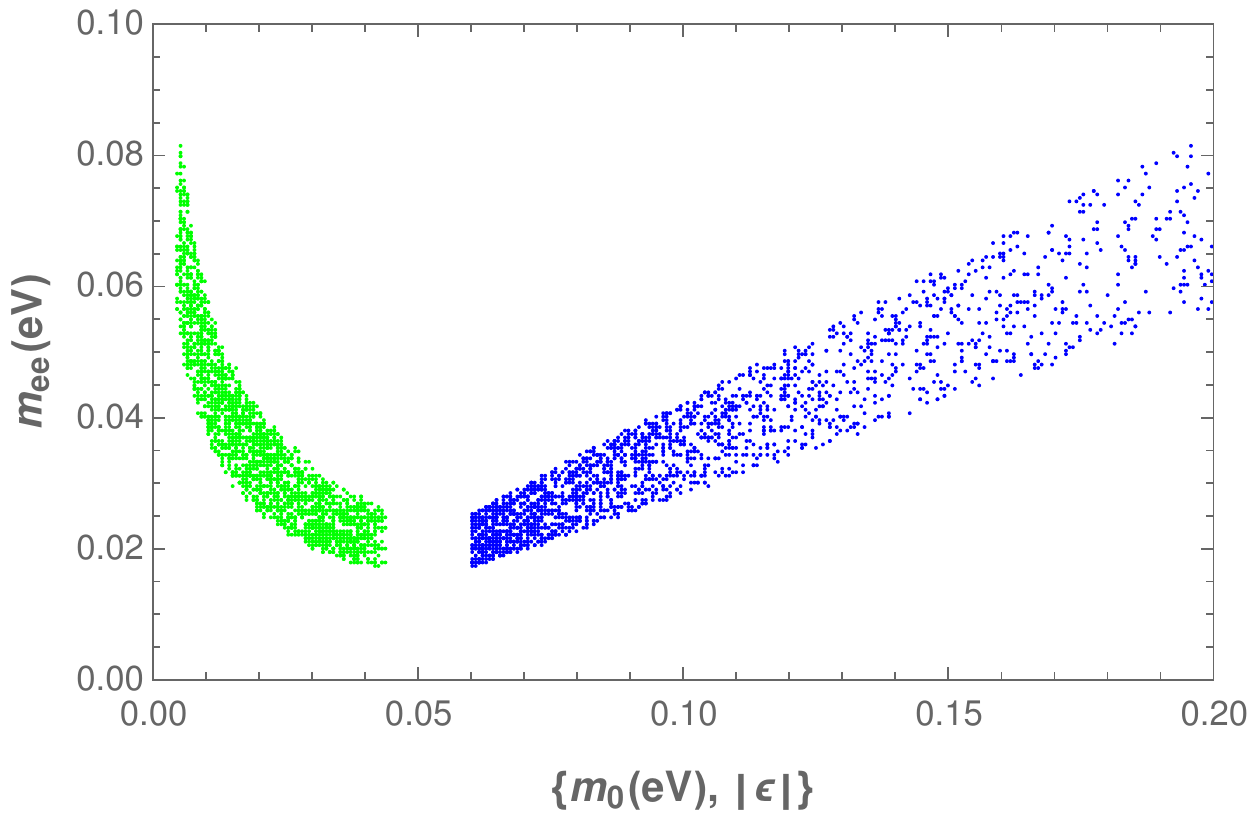}
\caption{From left to right:  the atmospheric angle versus (a) the
reactor angle , (b) $m_0$  and $\left|\epsilon
\right|$, and (c) $m_{ee}$ versus $m_0$  and $\left|\epsilon \right|$. Degenerate hierarchy: blue and green points stand for $m_{0}$ and $\left|\epsilon\right|$, respectively. The dotdashed, dashed and thick lines stand for $1~\sigma$, $2~\sigma$ and $3~\sigma$ of C. L.}\label{mee6a}
\end{figure}

In order to show that there is a parameter space for the $\epsilon$,
$\left|m^{0}_{\nu_{3}} \right|$ and $m_{0}$, we have made scattered
plots where we demand that the reactor and atmospheric angles lie within
$3\sigma$ of their experimental values, whereas the squared mass scales lie within $2\sigma$
\cite{deSalas:2017kay}. We allow $\left|\epsilon\right|$ and
$\left|m^{0}_{\nu_{3}}\right|$ ($m_{0}$) to vary from $0-0.3$ and
$0-0.1~eV$ ($0.06-0.2~eV$), respectively. Figures \ref{mee6} and
\ref{mee6a}  show the atmospheric angle versus the reactor angle in
panel (a),
and versus $\vert
m^{0}_{\nu_{3}}\vert$ ($m_{0}$) (in green) and $\left|\epsilon
\right|$ (in blue) in panel (b). At the
same time, as model prediction the effective neutrino mass rate for
neutrinoless double beta decay \cite{PhysRevD.73.053005, Rodejohann:2012xd, Bilenky:2012qi, Agostini:2013mzu} is displayed for inverted and degenerate
ordering in panel  (c).

\section{Conclusions}

An economical scalar extension of the $B-L$ gauge model has been built
for fermion masses and mixings. We have stressed that the very pronounced and the smaller hierarchy among the quark
and active neutrino masses respectively are the main motivations to
make an unusual assignment for the fermion families under the
${\bf S}_{3}$ discrete symmetry, that becomes fundamental to
understand the contrasting values between the CKM and PMNS mixing
matrices; the large hierarchy in the quark masses is reflected
in the hierarchical NNI textures that hijack the quark mass matrices,
and therefore, the CKM mixing. On the other hand, the lepton mixing might be explained by a
soft breaking of the $\mu \leftrightarrow \tau$ symmetry, where a set
of values for the relevant free parameters was found to be consistent
with the last experimental data on lepton observables.
The model also has a rich scalar sector, providing opportunities for
its experimental testing.

Last but not least, this naive work remarks that the non-abelian
group, ${\bf S}_{3}$, together with two ${\bf Z}_{2}$ parities may be
considered as the underlying flavor symmetry at low energies that
allows us to understand the fermion masses and mixings, even though
the lepton sector is limited in the sense that the Dirac CP violating
and Majorana phases are not predicted in the model.

\section*{Acknowledgements}
This work was partially supported by Mexican Grants 237004, PAPIIT
grant IN111518. I thank the Department of Theoretical Physics at IFUNAM for the warm
hospitality. I would like to make an especial mention to Gabriela
Nabor, Marisol, Cecilia and Elizabeth G\'omez for their financial and
moral support during this long time. We are in debt with our families.

\bibliographystyle{unsrt}
\bibliography{references.bib}
\end{document}